\documentclass[12pt,a4paper]{article}

\usepackage{bm}
\usepackage{ulem}
\usepackage[dvipsnames,usenames]{color}
\usepackage{amsmath,amssymb,latexsym}
\usepackage{graphicx}

\usepackage[numbers,sort&compress]{natbib}
\usepackage[labelfont={it},labelsep=period,font={small,it},justification=justified]{caption}

\usepackage{times}
\usepackage{mathptmx}

\usepackage{geometry}
\usepackage{fancyhdr}
\usepackage[tiny]{titlesec}
\titlelabel{\thetitle.\ }

\usepackage{setspace}
\newcommand{\PP}{{Phys.\ Plasmas}}

\begin{document}

\begin{doublespace}
\centering
{\bfseries\Large{Statistical description of turbulent transport for flux driven toroidal plasmas}}\\
\end{doublespace}
\mbox{} \\
\mbox{} \\
J.~Anderson 1), K.~Imadera 2), Y.~Kishimoto 2), J.Q.~Li 3), H.~Nordman 1)
\\ \mbox{} \\
1) Chalmers University of Technology, SE-412 96 G\"{o}teborg, Sweden\\
2) Department of Fundamental Energy Science, Graduate School of Energy Science, Kyoto University, Gokasho, Uji, Kyoto 611-0011, Japan 
3) Southwestern Institute of Physics, Chengdu, Sichuan 610041, China
\\ \mbox{} \\
E-mail contact of main author: anderson.johan@gmail.com\\
\mbox{}\\
\begin{abstract}
A novel methodology to analyze non-Gaussian probability distribution functions (PDFs) of intermittent turbulent transport in global full-f gyrokinetic simulations is presented. In this work, the Auto-Regressive Integrated Moving Average (ARIMA) model is applied to time series data of intermittent turbulent heat transport to separate noise and oscillatory trends, allowing for the extraction of non-Gaussian features of the PDFs. It was shown that non-Gaussian tails of the PDFs from first principles based gyrokinetic simulations agree with an analytical estimation based on a two fluid model.
\end{abstract}
\noindent
\section{\label{sec:level1} Introduction}
During recent years an overwhelming body of evidence shows that the overall transport of heat and particles is to a large part caused by intermittency (or bursty events) related to coherent structures\cite{zweben2007, politzer2000, beyer2000, drake1988, antar2001, carreras1996, nagashima2011}. One of the main challenges in magnetic fusion research has been to predict the turbulent heat and particle transport originating from various micro-instabilities. The ion-temperature-gradient (ITG) mode is one of the main candidates for causing the anomalous heat transport in core plasmas of tokamaks. A significant part of the anomalous heat transport can be mediated by coherent structures such as streamers and blobs through the formation of avalanche like events of large amplitude, as indicated by recent numerical studies. These events cause a deviation of the probability distribution functions (PDFs) from a Gaussian profile on which the traditional mean field theory (such as transport coefficients) is based.

A crucial question in plasma confinement is thus the prediction of the PDFs of the transport due to these structures and of their formation. This work provides a theoretical interpretation of numerically generated PDFs of intermittent plasma transport events as well as offering an explanation for elevated PDF tails of heat flux. Specifically, in this work we analyze time traces of heat flux generated by global nonlinear gyrokinetic simulations of ion-temperature-gradient turbulence by the GKNET software \cite{Imadera2014}. The simulation framework is global, flux-driven and considers adiabatic electrons. In these simulations SOC type intermittent bursts are frequently observed and transport is often regulated by non-diffusive processes, thus the PDFs of e.g. heat flux are in general non-Gaussian with enhanced tails. 

\section{\label{sec:level2} Statistical analysis}
We have performed a statistical analysis of a global GK simulation of ion-temperature-gradient mode turbulence with adiabatic electrons by the software Gyro-Kinetic based Numerical Experimental Tokamak (GKNET). In the statistical analysis PDFs of the original time traces are computed along with the modelled time traces using the mathematical AutoRegressive Integrated Moving Average (ARIMA). In this case, the time evolution of the ion heat flux is considered as a time series, to which we apply standard Box-Jenkins (ARIMA) modelling~\cite{box1994}. This mathematical procedure effectively removes deterministic autocorrelations from the system, allowing for the statistical interpretation of the residual part, which a posteriori turns out to be relevant for comparison with the analytical theory. In our setup, it turns out that an ARIMA(2,1,0) model accurately describes the stochastic procedure, in that, one can express the (differenced) heat flux time trace in the form
\begin{eqnarray}
Q_{t+\Delta t}=a_1\,Q_{t}+a_2\,Q_{t-\Delta t}+Q_{res}(t)
\end{eqnarray}
where the fitted coefficients $a_1, a_2$ describe the deterministic component and $Q_{res}(t)$ is the residual part (noise).

Using this analysis, it is systematically observed that the residual PDFs are manifestly non-Gaussian with elevated tails, as shown for instance in Fig.~\ref{fig:fig1}, where the sample residuals from a GKNET simulation are tested against the Gaussian distribution.

In the following paragraph, we briefly outline the implementation of the instanton method. For more details, the reader is referred to the existing literature ~\cite{justin1989, gurarie1996, falcovich1996, kim2002, kim2008, anderson1, Anderson2010, Anderson2014, Anderson2015}. The PDF tail is first formally expressed in terms of a path integral by utilizing the Gaussian statistics of the forcing in the continuity equation in a similar spirit as in~\cite{carreras1996}. An optimum path will then be associated with the creation of a modon (among all possible paths or functional values) and the action ($S_{\lambda}$, below) is evaluated using the saddle-point method on the effective action in the limit $\lambda \rightarrow \infty$. The optimum path is determined by those functions that optimises the action $S_{\lambda}$ The instanton is localized in time, existing during the formation of the modon. The saddle-point solution of the dynamical variable $\phi(\vec{x},t)$ of the form $\phi(\vec{x},t) = F(t) \psi(\vec{x})$ is called an instanton if $F(t) = 0$ at $t=-\infty$ and $F(t) \neq 0$ at $t=0$. Note that, the function $\psi(\vec{x})$ here represents the spatial form of the coherent structure. Thus, the intermittent character of the transport consisting of bursty events can be described by the creation of modons. The probability density function of the heat flux $Q$ can be defined as 
\begin{eqnarray} \label{pq}
P(Q) =  \langle \delta(v_r n T_i(\vec{r}=\vec{x}_0)) - Q) \rangle = \int d \lambda e^{i \lambda Q} I_{\lambda},
\end{eqnarray}
where 
\begin{eqnarray} \label{ilambda1}
I_{\lambda} = \langle \exp(-i \lambda v_r n T_i(\vec{x}=\vec{x}_0)) \rangle.
\end{eqnarray}
Here, $v_{r}$ is the radial drift velocity and $T_i$ the ion temperature. The integrand can then be rewritten in the form of a path-integral as
\begin{eqnarray} \label{ilambda2}
I_{\lambda} = \int \mathcal{D} \phi \mathcal{D} \bar{\phi} e^{-S_{\lambda}}.
\end{eqnarray}
Although $\bar{\phi}$ appears to be simply a convenient mathematical tool, it does have a useful physical meaning that should be noted; it arises from the uncertainty in the value of $\phi$ due to the stochastic forcing. That is, the dynamical system with a stochastic forcing should be extended to a larger space involving this conjugate variable, whereby $\phi$ and $\bar{\phi}$ constitute an uncertainty relation. Furthermore, $\bar{\phi}$ acts as a mediator between the observables (heat flux) and instantons (physical variables) through stochastic forcing. In Eq.~\ref{ilambda2}, the integral in $\lambda$ is computed using the saddle-point method where it is shown that the limit $\lambda \rightarrow \infty$ corresponds to $Q \rightarrow \infty$, representing the tail part of the distribution. Based on the assumption that the total PDF can be characterized by an exponential form, the expression
\begin{eqnarray} \label{pq2}
P(Q) & = & \frac{1}{Nb} \exp{\{ - b |Q-\mu |^{3/2}\}}, \\ \label{b}
b & = & b_0 (\frac{R}{L_n} + 2 \langle g_i \rangle \beta - U - \langle k_{\perp}^2\rangle (U + \frac{R}{L_n})), \\
\beta & = & 2 + \frac{2}{3} \frac{R/L_n - U}{U + 10/3 \tau \langle g_i \rangle}.
\end{eqnarray}
is found, where the heat flux $Q$ plays the role of the stochastic variable, with $P(Q)$ determining its statistical properties. Several auxiliary definitions are also utilized; the normalization constant $N$; the gradient scale lengths $L_f = - \left( d ln f / dr\right)^{-1}$; the normalized modon speed ${U} = R U/L_n$ and temperature ratio $\tau = T_i/T_e$; $g_i = \omega_d/\omega_{\star} = \frac{2 L_n}{R} (\cos (\xi) + \hat{s} \xi \sin(\xi))$ where $\omega_d$ is the curvature drift frequency and $\omega_{\star}$ is the diamagnetic drift frequency; $k_{\perp}^{2} = k_y^2 (1 + \hat{s}^2 \xi^2)$ is the perpendicular wave number; the brackets denote averaging along the field line, e.g. for an arbitrary scalar function $f$, $\langle f \rangle = \int_{- \pi}^{\pi} d\xi \phi f \phi/ \int_{-\pi}^{\pi} d \xi \phi^2$ where the eigenfunctions $\phi (\xi)$ have to assumed or taken from a simulation. The coefficient $b_0$ is a free parameter and represents the strength of the forcing in the continuity equation. Note that the proposed PDF is close enough to a Gaussian distribution to match the bulk of the PDF while retaining the enhanced tails. Furthermore, since the physical constants are available matching of the PDF over a large range of parameters is possible unlike matching each case with a Gaussian distribution.

\section{\label{sec:level3} Results}
As a simulation framework we use a circular concentric tokamak configuration with $R_0/a = 2.79$, $a/\rho_{ti} = 150$  and $q(r) = 0.85 + 2.18 (\frac{r}{a})^2$ . Initial plasma parameters at $r/a = 0.5$  are  $(R_0/L_n)_{r=a_0/2} = 2.22$, $(R_0/L_{Ti})_{r=a_0/2} = 10.0$, $(R_0/L_{Te})_{r=a_0/2} = 6.92$, and  $\nu_{\star} = 0.28$, respectively. In this configuration, simulation parameters are taken as follows; the time step length is $\Delta t = 2 \times 10^{-3} R_0/v_{ti}$, grid number and system size are $(N_r, N_{\theta}, N_{\phi}, N_{v_{||}}, N_{\mu}) = (128, 128, 64, 64, 16)$ and $(L_r, L_{\theta}, L_{\phi}, L_{v_{||}}, L_{\mu}) = (150 \rho_i, 2 \pi, \pi, 12 v_{ti}, 18 v_{ti}^2/B_0)$, respectively. Note that a $1/2$ wedge torus is employed in this simulation. Figure 1 shows (a) initial density, temperature and safety factor profiles, and (b) deposition profiles of   $A_{src}$ and $A_{snk}$. Here, the source and sink parameters are chosen to avoid large deviations from the Maxwellian distribution in the heating region and possible un-physical oscillations trigged by the fixed outer boundary condition. Within this framework, we perform gyrokinetic simulations of flux-driven toroidal ITG turbulence with external heat input $P_{in} = 16$ [MW]. Note that not only turbulence and zonal flow, but also the neoclassical transport and mean flow determined self-consistently by evolving equilibrium profiles can be properly traced in this framework.
Based on this simulation set-up we have performed a statistical analysis of one base case where the parameters are similar to the cyclone base case parameters, i.e. $a/\rho_i = 150$, $a/R_0 = 0.36$ and $\tau_{snk}^{-1} R_0/v_{ti} = 0.25$. 

We first present the PDFs \ref{fig:fig1} of the original time series at certain locations in $r/\rho_i$ (30, 42, 54, 66, 77, 89, 100, 112, 124, 148). In order to be able to perform an ARIMA analysis with high confidence a reasonably large data set is needed. Note that the PDFs at each location each radial position (128 in total) are analysed with 4000 data points in each time trace. In particular, this is needed to capture the higher statistical moments such as skewness and kurtosis.

The PDFs of the corresponding ARIMA modelled time traces are presented in Figure (\ref{fig:fig1}) and contrasted to Gaussian distributions and the analytical prediction found in Eq. (\ref{pq2}) at certain radial positions. The number of PDFs are restricted due to easier identification of PDF and radial position. This is corroborating results found previously in comparing analytical modelling and local GK results using {\sc GENE} \cite{Anderson2010} for the right tail whereas the left tail falls off faster then the predicted (right figure). Note that, in the analytical model only the right tail was computed and it was then assumed to be quasi-symmetric. We find in the left figure (\ref{fig:fig1}) that the PDFs significantly change form as we move towards the edge. Furthermore, small but finite asymmetry can be observed with slightly larger tails on the right compared to the left side. This would indicate a net heat flux moving outwards. To quantify this we compute the skewness (third moment) of the time traces. Note that elevated tails is a signature of intermittency and transport mediated by large structures. 

The kurtosis (left) and the skewness (right) are computed of the original and the ARIMA modelled time traces as a function of radial position are presented in Figure (\ref{fig:fig2}). Note that a Gaussian distribution has a kurtosis exactly equal to 3. It is found that for many radial positions that the kurtosis (left figure) is above 3 which indicate super-diffusive transport often mediated by large structures such as streamers of blobs. For the skewness (right figure) it is identified that in most radial positions the skewness is positive indicating that a positive heat flux transport outwards is slightly more likely than the opposite giving a net heat flux outwards.

\section*{Shear scan}
In order to investigate different magnetic field configurations we have investigated two different safety factor ($q$) profiles yielding widely different magnetic shear variations. Note that the input power is on 4MW compared to the 16MW in the base case, i.e. $a/R_0 = 0.36$, $(R_0/L_n)_{r=a_0/2} = 2.22$, $(R_0/L_{Ti})_{r=a_0/2} = 10.0$, $(R_0/L_{Te})_{r=a_0/2} = 6.92$, $\nu_{\star} = 0.28$, $P = 16 MW$, $\tau_{snk}^{-1} R_0/v_{ti} = 0.25$. 
The safety factor profile is given by the expression for $n = 2$, $4$ and $6$,
\begin{equation}
q_n(r) = 0.85 + 2.18(r/a)^n
\end{equation}
which also determines the shear profiles.

The resulting PDFs of of the heat flux time traces generated using the two safety factor profiles, 2nd and 6th order, are presented in Figure \ref{f:PDFshear}. Note that there is a difference in the total simulation time and length of the time traces. In the 2nd order case (left figure) the total simulation time is 800 (4000 simulation time points) whereas in the 6th order  (right figure)simulation it is 2400 (12000 simulation time points). There are some notable differences in the in two PDFs, it is indicated that in the 6th order safety factor profile there is a higher likelyhood for smaller heat flux events whereas the PDF of the 2nd order safety factor profile is significantly broader and thus allows for higher likelyhood of larger heat flux events. 

Next we present the PDFs from the ARIMA modelled time traces in both 2nd (left figure) and 6th (right figure) order cases. We find that some of the PDFs stemming from time traces generated in 2nd order  safety factor profile case are significantly broader and thus allows for high heat flux events or transport by coherent structures. We note that all PDFs are non-Gaussian. Here it is likely that there are some effects of the differences in the length of the time traces, in particular it seems that the PDFs at the edge are not fully converged for the 2nd order safety factor profile.

In Figure \ref{f:PDFshear2} it is indicated that there exist a transition in the type of transport along the radius which reflects in that the tails of the PDFs changes. Thus we have made an overfitting with two different distributions at two diffent locations in Figure \ref{f:PDFshear_f}. For source and intermediate regions we find a centrally smoother PDF with elevated tails compared to a Gaussian and at the edge we find a clearly Laplacian type of PDF with a cusp at the center. The PDFs in the source and intermediate regions are reasonably well fitted by the analytical model and the edge PDFs are asymmetrically possible to fit with a Laplacian distribution.

One possible way of checking the resolution and how well the ARIMA model can represent the original time trace is by comparing the kurtosis (4th moment) since this should be approximately preserved comparing the original time trace and the ARIMA modelled counterpart. We find a significantly better match in using the longer time trace of the 6th order safety factor profile which is three times longer compared to a shorter time trace. This is due to collecting rare events that produces deviations from Gaussianity in the PDFs is exceedingly time consuming, furthermore it seems to be the general case. However there are a few radial points where we find differences. In particular, we note that for the edge it seems that the PDFs are not completely converged even for the longer time series for the 6th order case. AS is displayed by deviations between the model and the simulation kurtosis profiles.

\section{\label{sec:level10} Discussion and Conclusion}
Here it is important to note that the gradients are not fixed during the simulation and e.g. that $R/L_{Ti}$ will exhibit a PDF with mean around 6-8 varying along the radius. The numerically generated time traces are processed with Box – Jenkins modelling in order to remove deterministic autocorrelations, thus retaining their stochastic parts only, a comparison with the original PDFs are shown in Figure \ref{fig:fig1}. The accuracy of the modelling can be evaluated by comparing the higher even statistical moments (kurtosis) of the PDFs, here a quite good representation of the kurtosis in the modelling is found as displayed in Figure \ref{fig:fig2}. Although, the same PDFs were previously found in local gyrokinetic simulations \cite{Anderson2010} there are some unique features present inherently coming from the global nature of the physics. We consistently find non-Gaussian distributions at most radial positions allowing for significant transport mediated by coherent structures such as blobs/coherent structures or streamers/avalanches, see Figure \ref{fig:fig2}. In the simulations large avalanche like structures are present as an indication of this mode of transport. In the statistical analysis the length of the simulations is of great importance and it is taken to be 800 (4000 time steps) although we have analyzed a three times longer time trace with similar results. The analytical model approximately reproduces the PDFs at various locations as is shown in Figure \ref{fig:fig2}. In the case of n=6 safety factor profile there is a change in the PDFs as we go to the edge where the PDFs are Laplacian distributed, supposedly this is due to the change from a flatter to steeper profile compared to the lower n cases. Furthermore, we note that a positive skewness indicates excess transport outwards in radial direction.
The main part of this work consists in providing a theoretical interpretation of the PDFs of radial heat flux derived by nonlinear, global, gyrokinetic simulations of drift-wave turbulence in tokamaks. These PDFs have been shown to agree very well with analytical predictions based on a fluid model, on applying the instanton method. The result points to a universality in the modelling of intermittent stochastic process while the analytical theory offers predictive capability, extending the previous result to be globally applicable.

We have presented a first quantitative comparison between a first-principles theoretical model of drift wave turbulence with self-consistent non-linear simulations. A key finding of this work is that the intermittent process in the context of drift-wave turbulence appears to be independent of the specific modelling framework, opening the way to the prediction of its salient features. The methodology has been validated using various different modelling strategies and simulation softwares \cite{Anderson2014, Anderson2015}. Specifically, we were able to quantitatively confirm the exponential form of the PDFs, therefore adding the important element of predictive strength to the existing phenomenological approaches. More importantly, a strong indication of universality in the description of drift wave turbulence has emerged.

\begin{figure}[ht]
{\vspace{2mm}
\includegraphics[width=14cm, height=14cm]{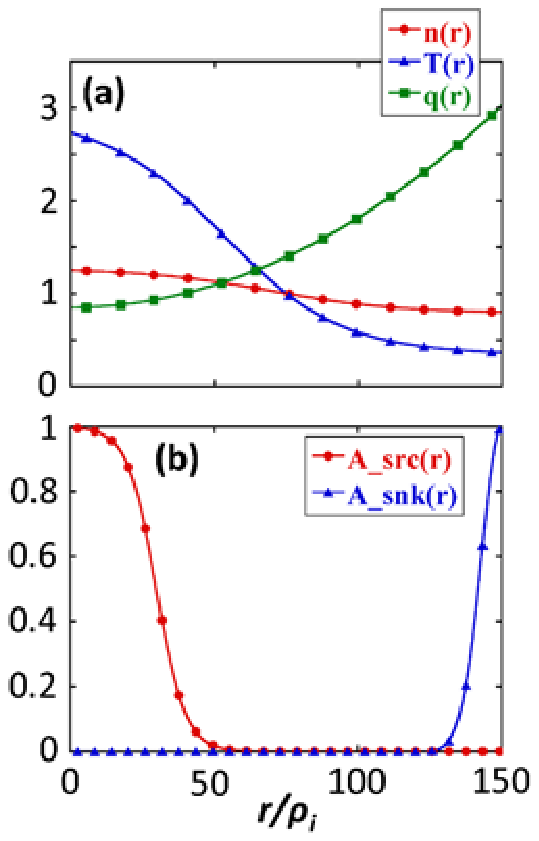}}
\caption{
(a) Initial density, temperature and safety factor profiles. (b) Deposition profiles of $A_{src}$ and $A_{snk}$.
}
\label{f:sim_setup_fig}
\end{figure}

\begin{figure}[ht!]
\begin{minipage}[b]{0.5\linewidth}
\centering
\includegraphics[width=9cm, height = 9cm]{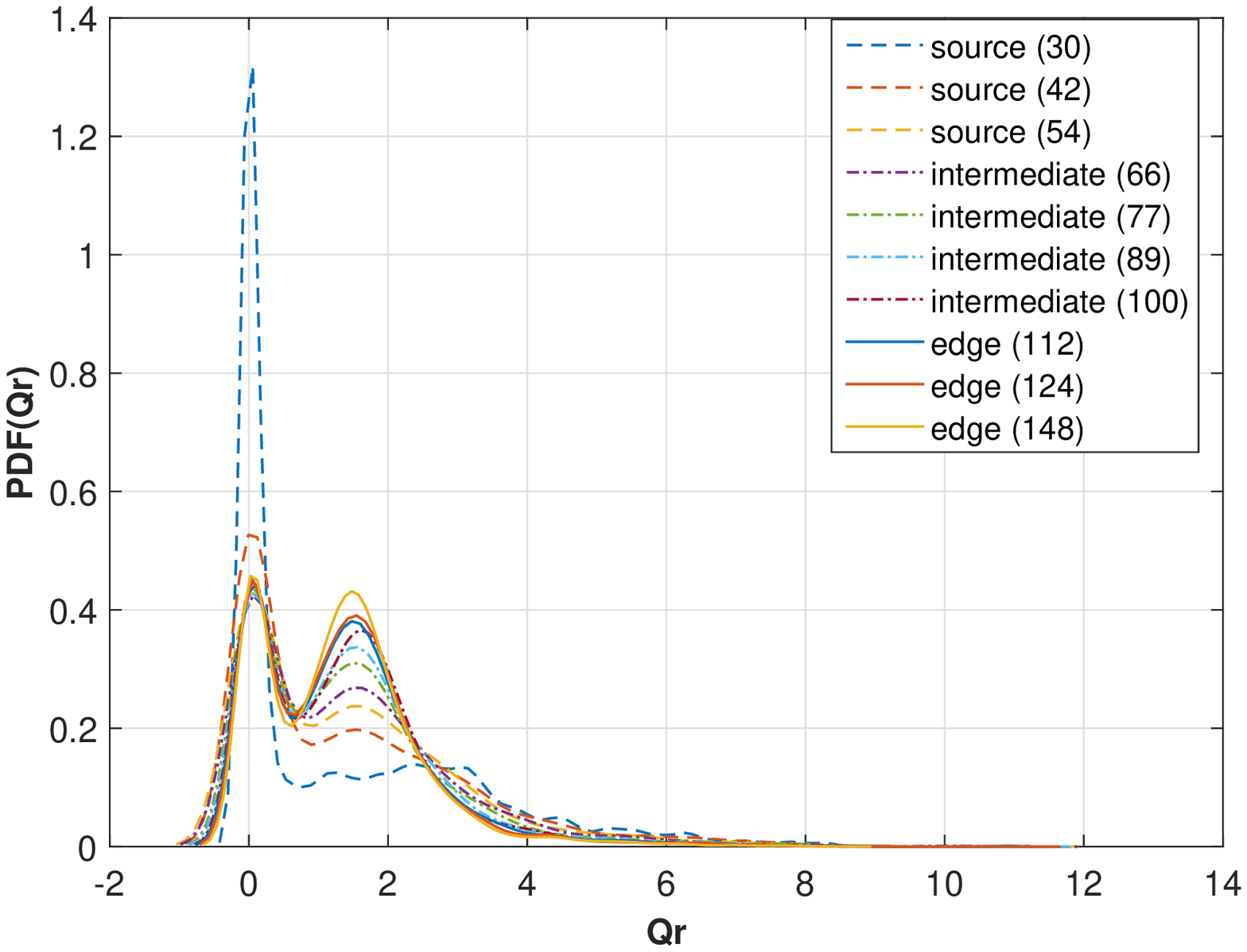}
\end{minipage}
\hspace{0.5cm}
\begin{minipage}[b]{0.5\linewidth}
\centering
\includegraphics[width=9cm, height = 9cm]{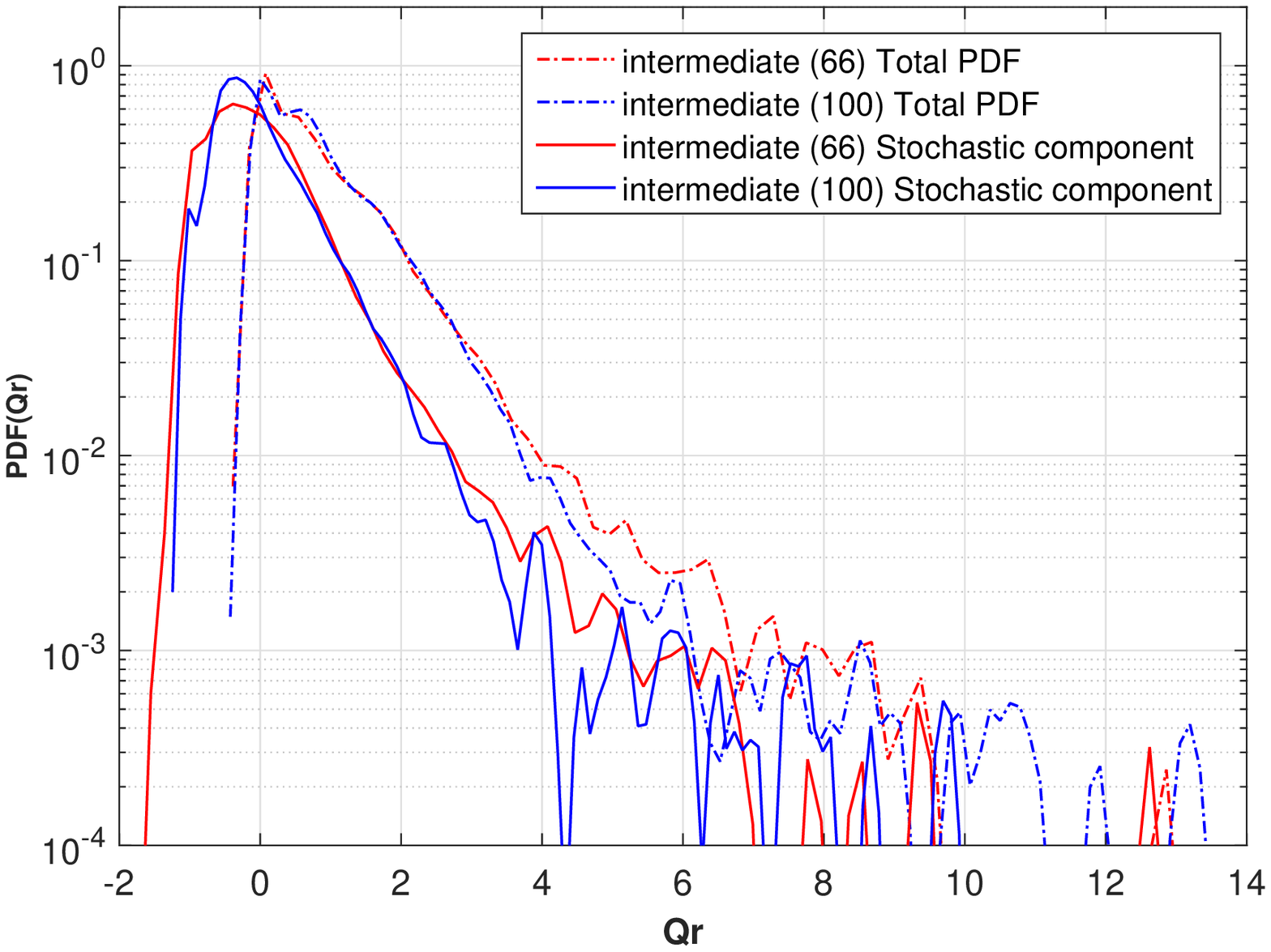}
\end{minipage}
\caption{(left) displays the PDFs at ten positions using Cyclone Base Case parameters and (right) shows the original PDFs and the model PDFs at two different locations.}
\label{fig:fig1}
\end{figure}

\begin{figure}[ht!]
\begin{minipage}[b]{0.5\linewidth}
\centering
\includegraphics[width=9cm, height = 9cm]{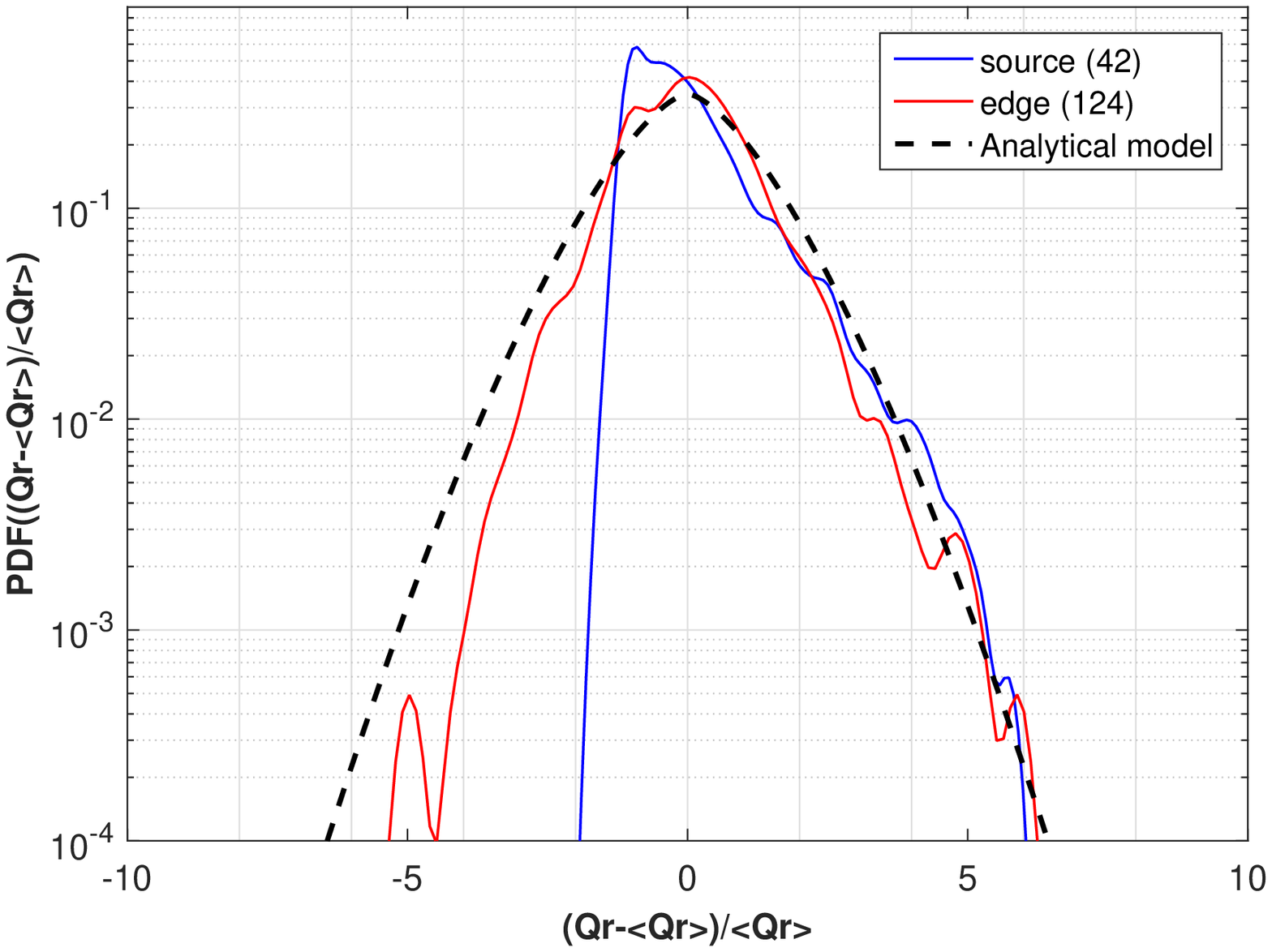}
\end{minipage}
\hspace{0.5cm}
\begin{minipage}[b]{0.5\linewidth}
\centering
\includegraphics[width=9cm, height = 9cm]{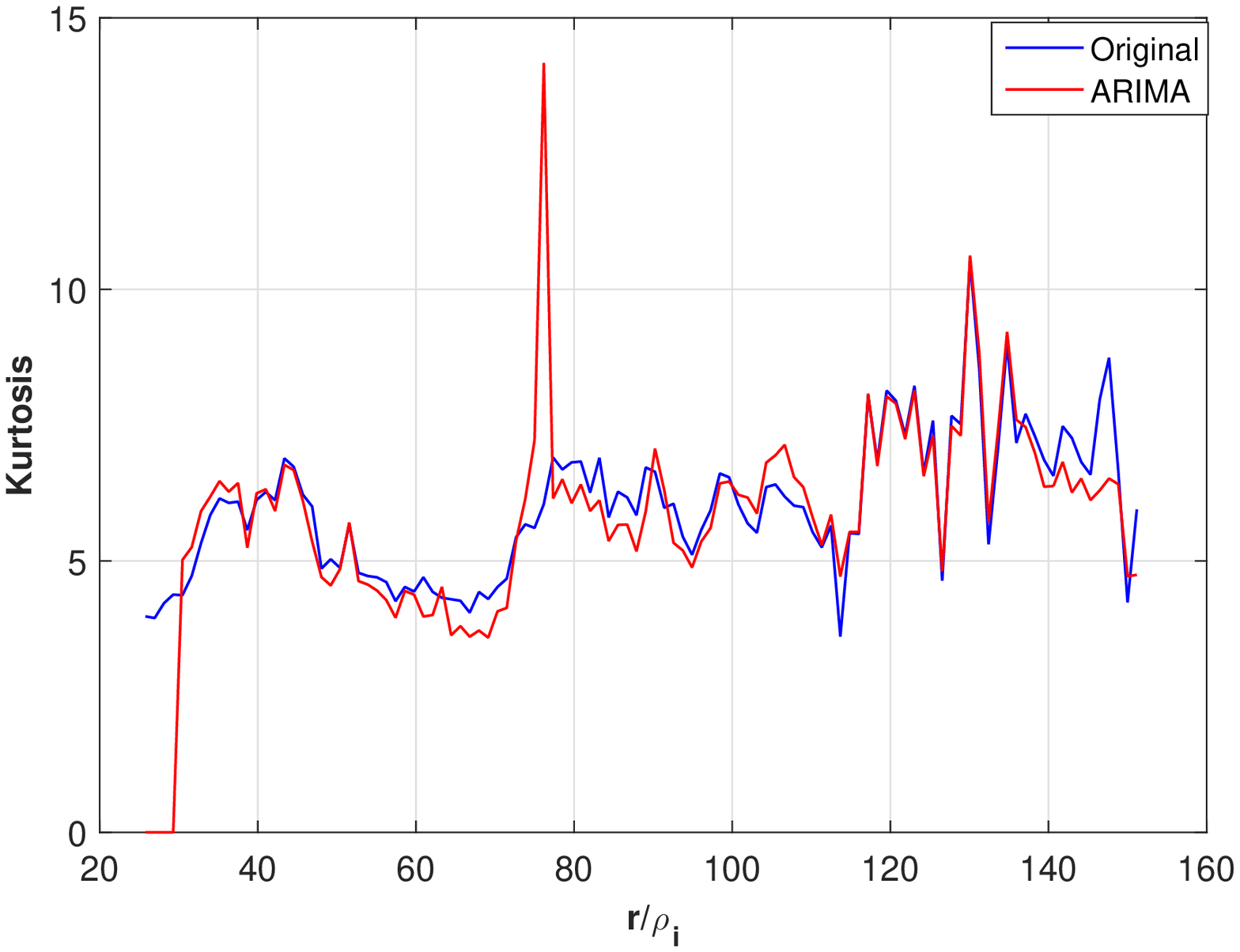}
\end{minipage}
\caption{(left) a comparison between the numerically found PDFs and the analytical model (n=2) and (right) a comparison of the kurtosis of the original and the stochastic model (n=6).}
\label{fig:fig2}
\end{figure}

\begin{figure}[ht]
{\vspace{2mm}
\includegraphics[width=9cm, height=9cm]{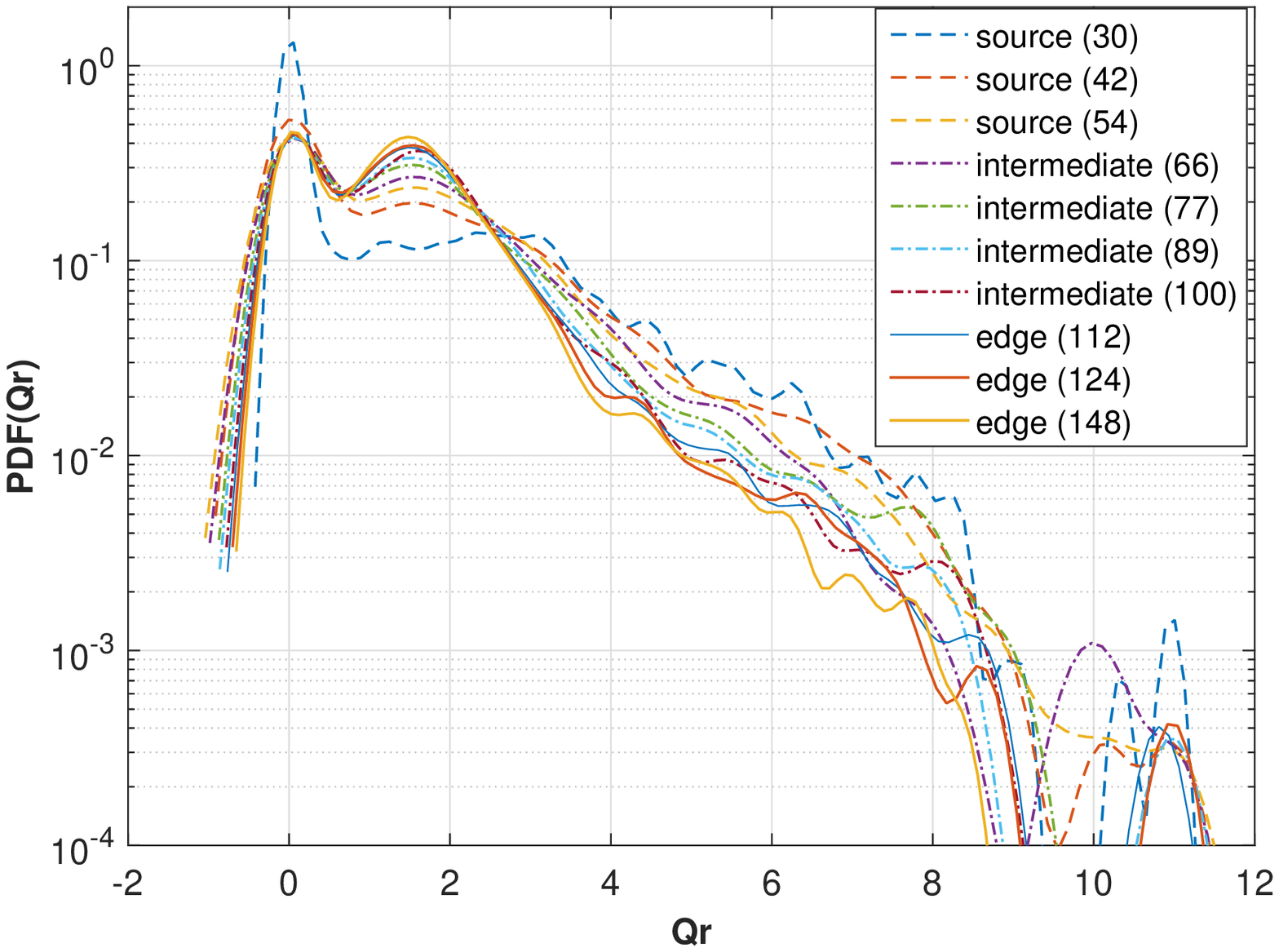}}
{\includegraphics[width=9cm, height=9cm]{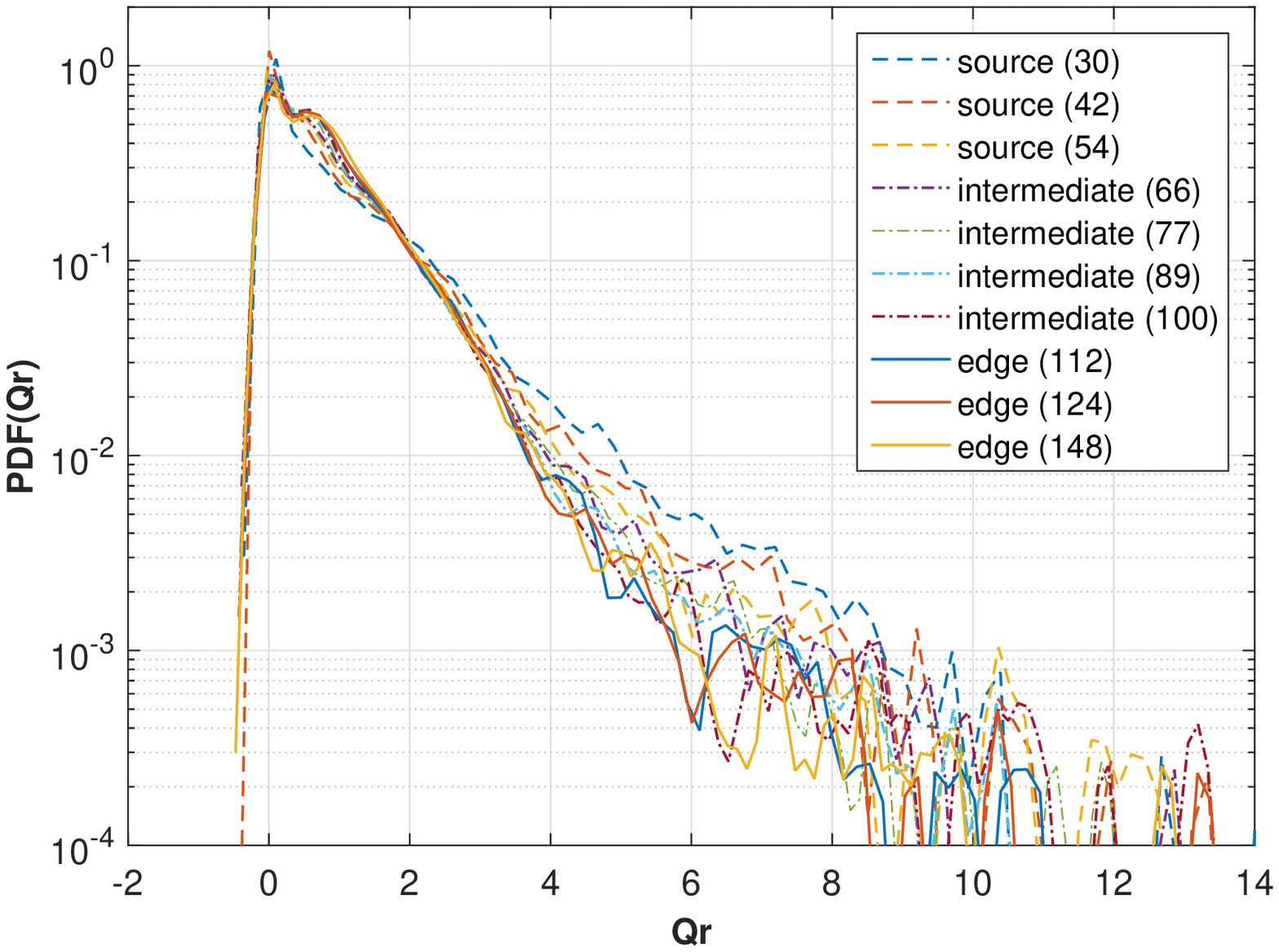}}
\caption{
The PDFs generated from the original time series at ten locations along the radius in log-scale. The second order safety factor profile is on the left and the 6th order profile to the right.
}
\label{f:PDFshear}
\end{figure}

\begin{figure}[ht]
{\vspace{2mm}
\includegraphics[width=9cm, height=9cm]{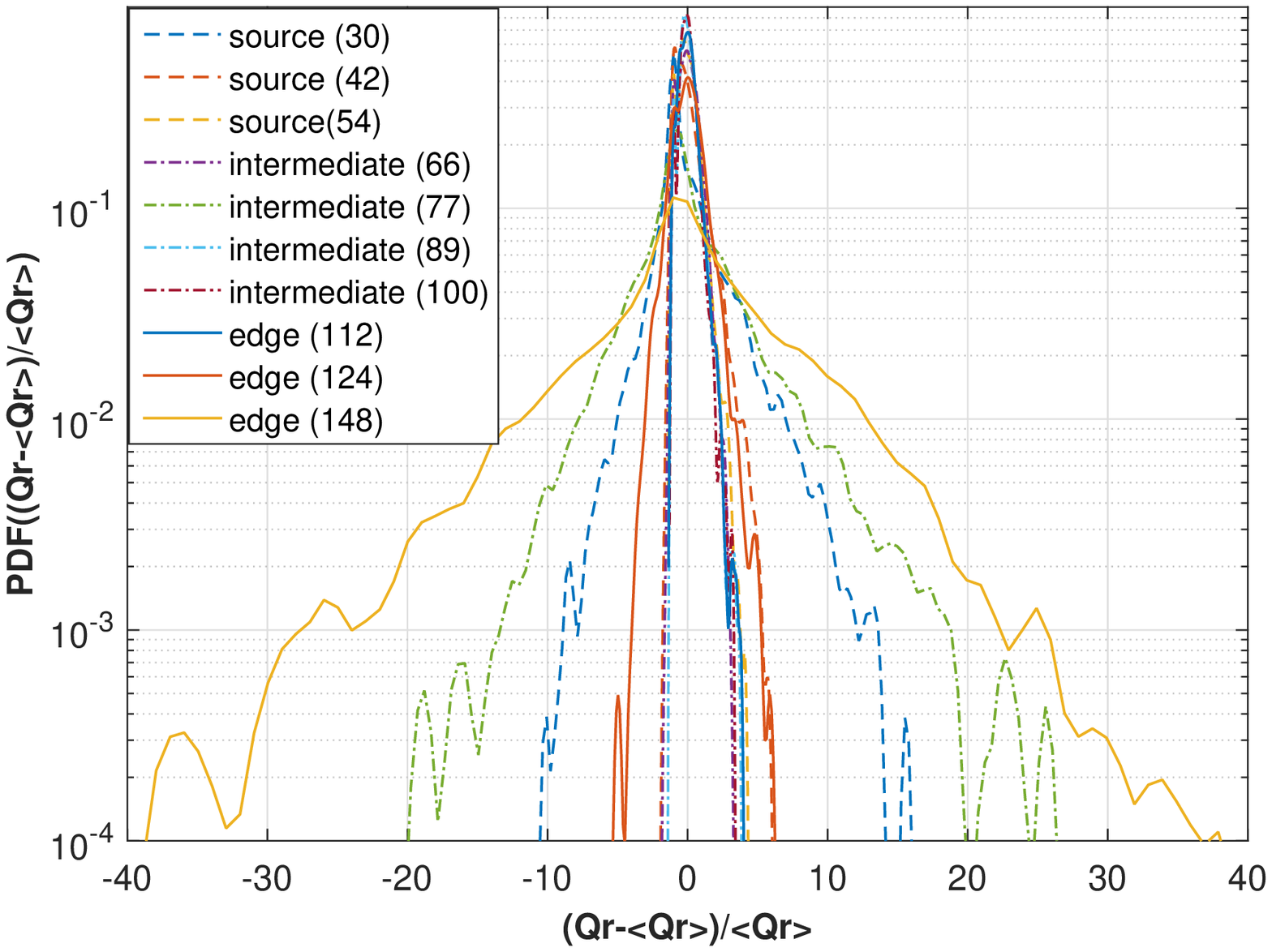}}
{\includegraphics[width=9cm, height=9cm]{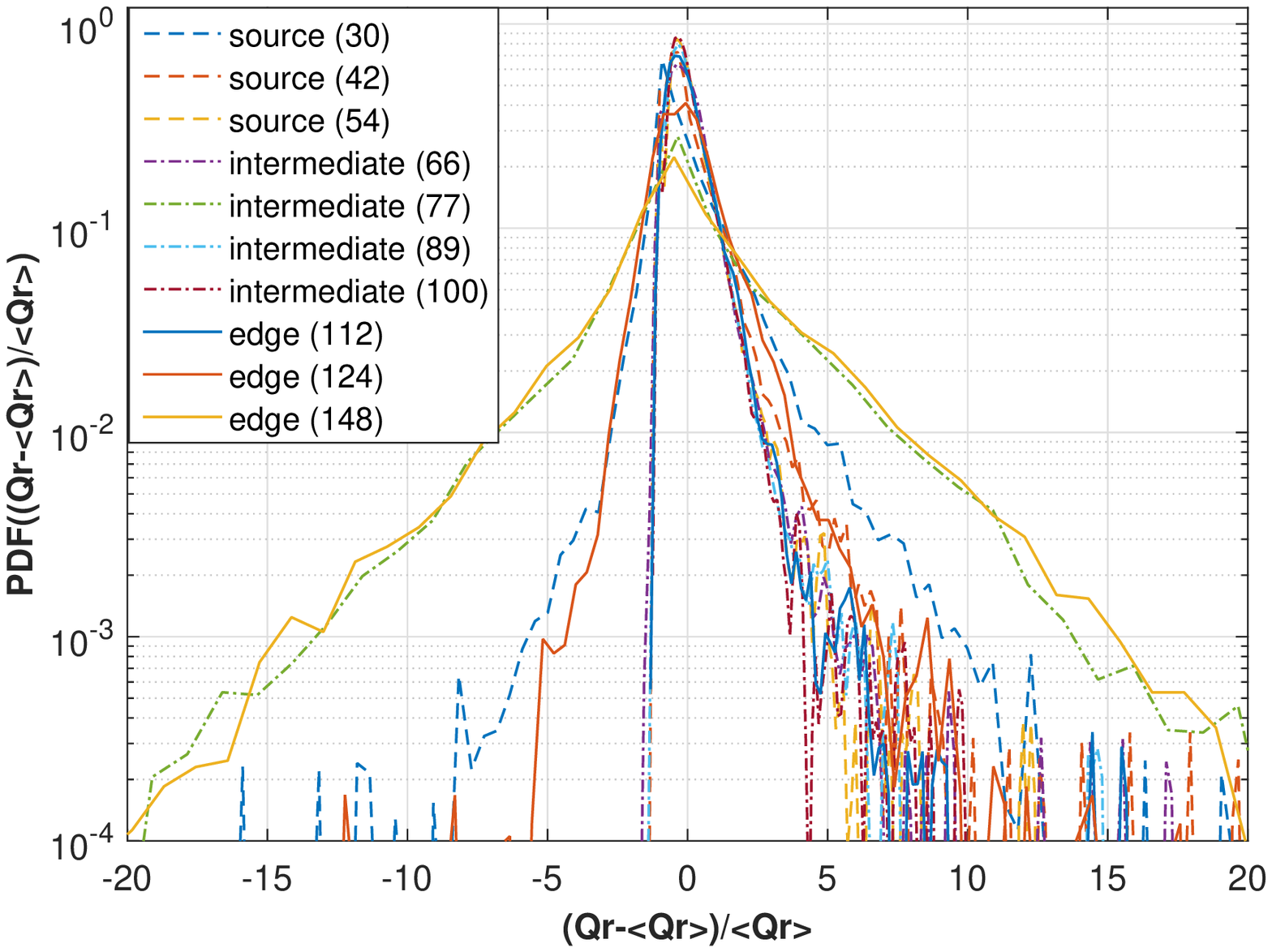}}
\caption{
The PDFs generated from the ARIMA modelled time traces at ten locations along the radius in log-scale. The 2nd order safety factor profile is on the left and the 6th order safety factor profile to the right.
}
\label{f:PDFshear2}
\end{figure}

\begin{figure}[ht]
{\vspace{2mm}
\includegraphics[width=9cm, height=9cm]{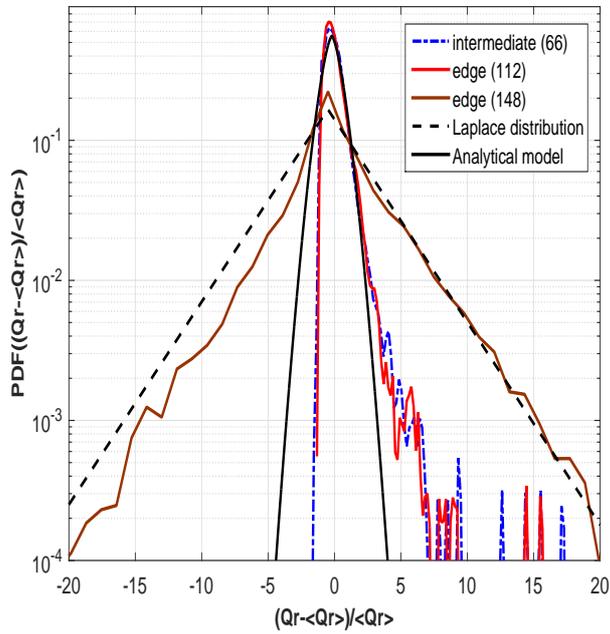}}
\caption{
Fits to the PDFs generated from the ARIMA modelled time traces at three locations along the radius for the 6th order safety factor profile.
}
\label{f:PDFshear_f}
\end{figure}

\begin{figure}[ht]
{\vspace{2mm}
\includegraphics[width=9cm, height=9cm]{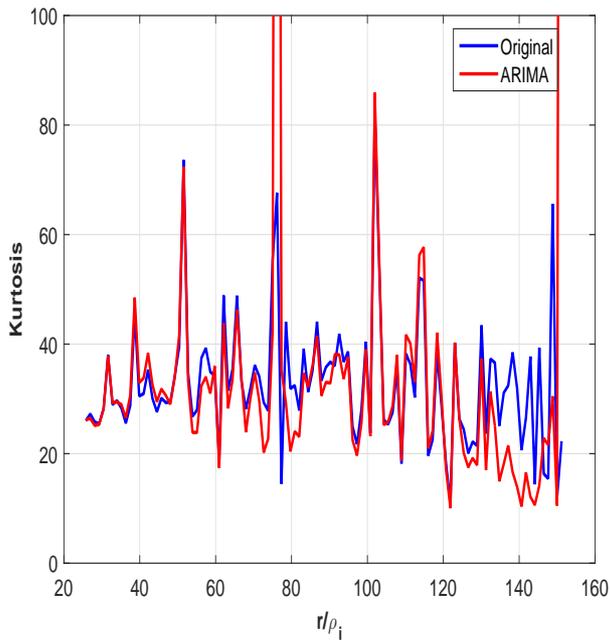}}
\caption{
The kurtosis of the time traces along the radius for 6th order safety factor profiles.
}
\label{f:PDFshear_k}
\end{figure}

\begin{figure}[ht]
{\vspace{2mm}
\includegraphics[width=9cm, height=9cm]{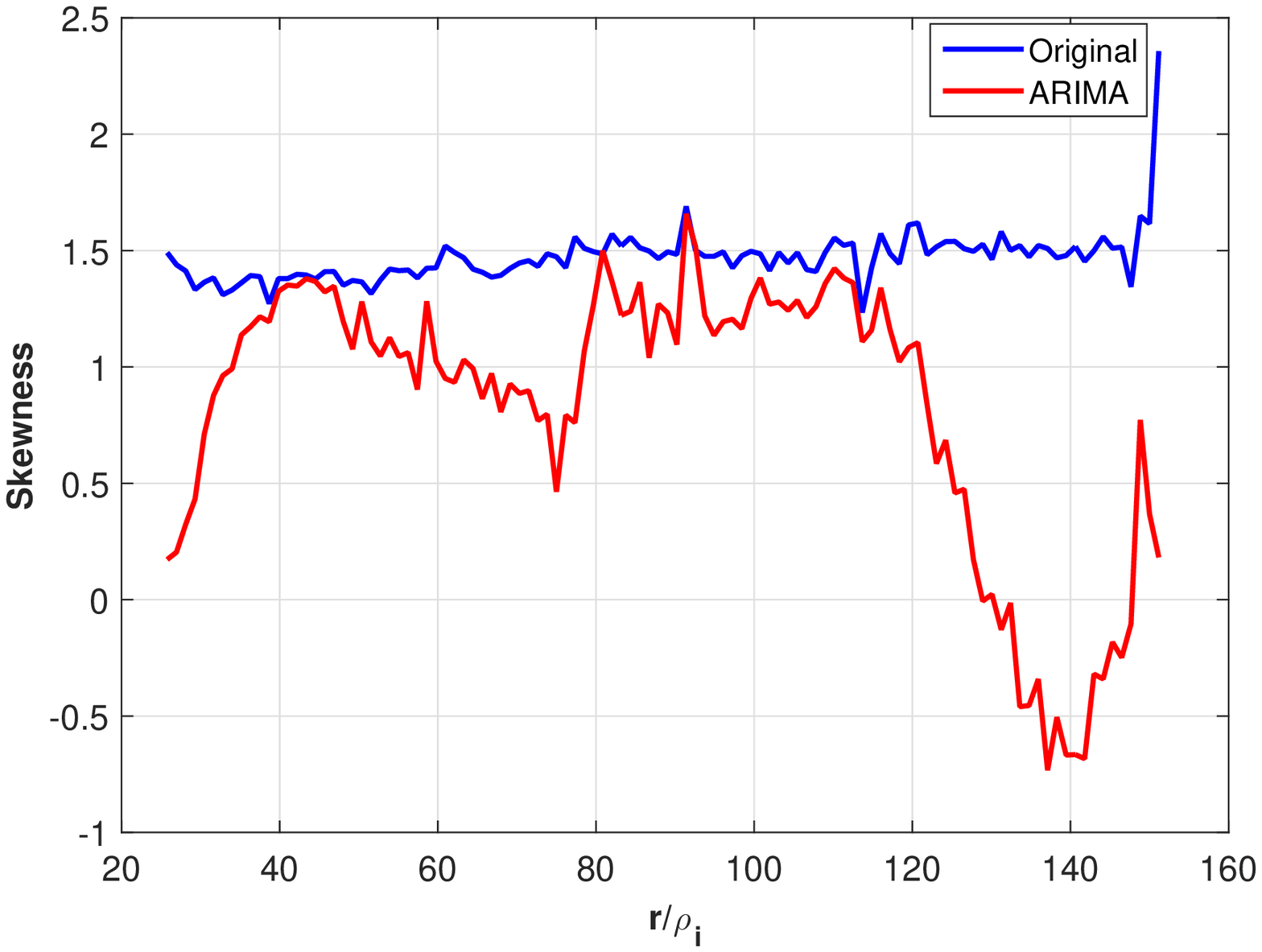}}
{\includegraphics[width=9cm, height=9cm]{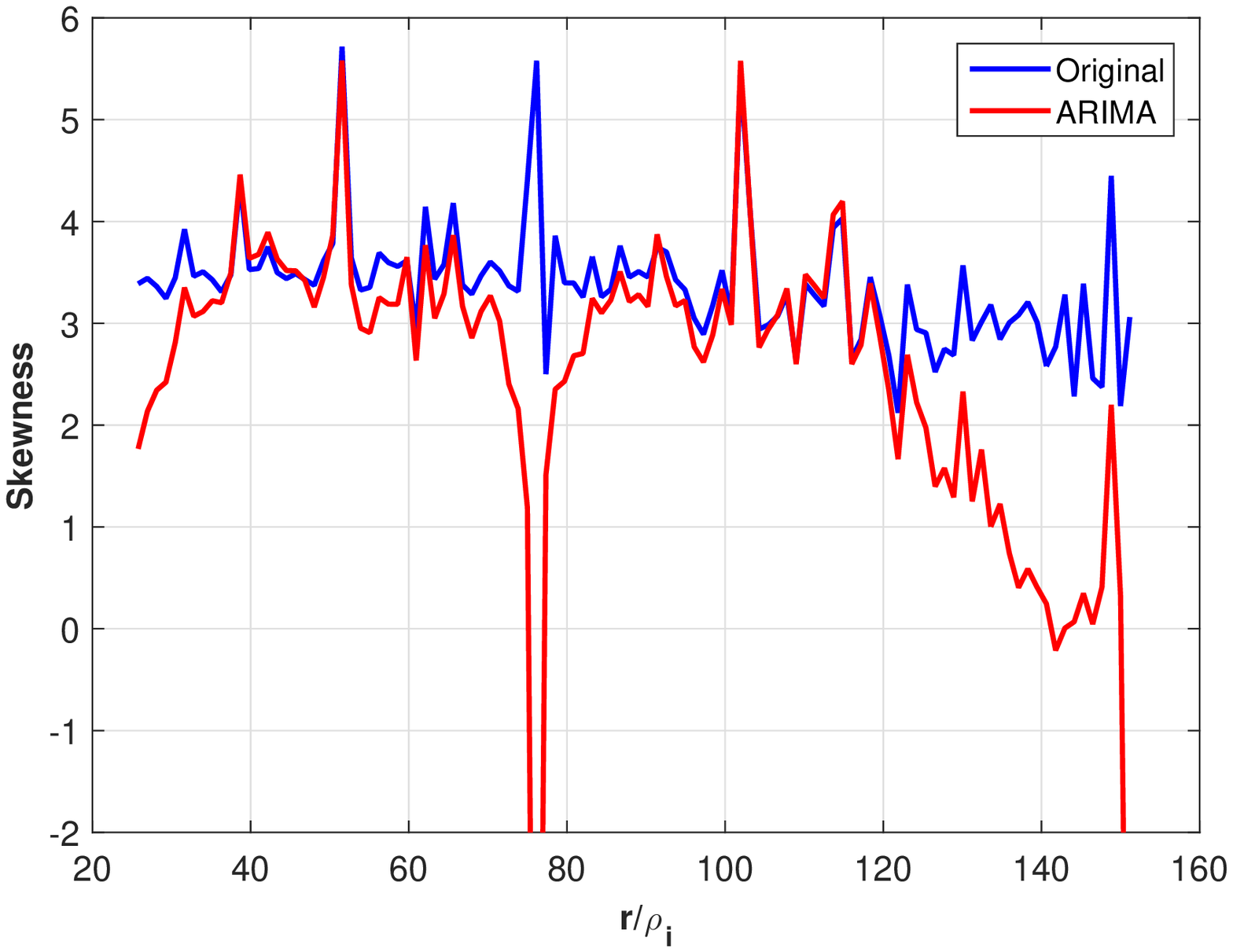}}
\caption{
The skewness of the time traces along the radius for the 2nd order (left) and 6th order (right) safety factor profiles.
}
\label{f:PDFshear_s}
\end{figure}

\end{document}